%% file: computation_paper.tex
\documentclass[12pt]{article}

\pdfoutput=1

\input{packages.tex}
\input{commands.tex}

\input{formatting.tex}

\begin{document}

\input{title.tex}
\bigskip

\begin{abstract}
\input{abstract.tex}
\end{abstract}
\noindent%
\vfill

\newpage
\spacingset{1.45}

\input{body.tex}

\input{acknowledgements.tex}

\bibliographystyle{agsm}
\bibliography{computation_paper}

\appendix
\input{appendix.tex}

\end{document}

%% file: packages.tex
\usepackage{amsfonts}
\usepackage{amsmath}
\usepackage{graphicx,psfrag,epsf}
\usepackage{enumerate}
\usepackage{natbib}
\usepackage{algorithm}
\usepackage{color}
\usepackage{setspace}
\usepackage{url} % not crucial - just used below for the URL
\usepackage{caption}
\usepackage{topcapt}

%% file: commands.tex
\providecommand{\blind}{0}
\providecommand{\fbseq}{\if0\blind{\tt fbseq}\fi \if1\blind{XX-BLIND-XX} \fi}
\providecommand{\fbseqCUDA}{\if0\blind{\tt fbseqCUDA}\fi \if1\blind{XX-BLIND-2-XX} \fi}
\providecommand{\fbseqComputation}{\if0\blind{\tt fbseqComputation}\fi \if1\blind{XX-BLIND-3-XX} \fi}
\providecommand{\RNAseq}{RNA-seq} 
\providecommand{\I}{\mathrm{I}}
\providecommand{\ind}{\stackrel{\text{ind}}{\sim}}
\providecommand{\ov}[1]{\overline{#1}}
\providecommand{\prl}{{\bf{In parallel,}}}
\providecommand{\red}{{\bf{Reduction}}}

\providecommand{\Unif}{\text{Uniform}}

%% file: formatting.tex
\def\spacingset#1{\renewcommand{\baselinestretch}%
{#1}\small\normalsize} \spacingset{1}

%% file: title.tex
% !TEX root = ../computation_paper.tex

\if0\blind
{
  \title{\bf A fully Bayesian strategy for high-dimensional hierarchical modeling using massively parallel computing}
  \author{Will Landau and Jarad Niemi\footnote{Contact the corresponding author at niemi@iastate.edu.}
  \hspace{.2cm}\\
    Department of Statistics, Iowa State University}
  \maketitle
} \fi

\if1\blind
{
  \bigskip
  \bigskip
  \bigskip
  \begin{center}
    {\LARGE\bf A fully Bayesian strategy for high-dimensional hierarchical modeling using massively parallel computing}
\end{center}
  \medskip
} \fi

%% file: abstract.tex
% !TEX root = ../computation_paper.tex

Markov chain Monte Carlo (MCMC) is the predominant tool used in Bayesian parameter estimation for hierarchical models. When the model expands due to an increasing number of hierarchical levels, number of groups at a particular level, or number of observations in each group, a fully Bayesian analysis via MCMC can easily become computationally demanding, even intractable. 
We illustrate how the steps in an MCMC for hierarchical models are predominantly one of two types: conditionally independent draws or low-dimensional draws based on summary statistics of parameters at higher levels of the hierarchy. 
Parallel computing can increase efficiency by performing embarrassingly parallel computations for conditionally independent draws and calculating the summary statistics using parallel reductions. 
During the MCMC algorithm, we record running means and means of squared parameter values to allow convergence diagnosis and posterior inference while avoiding the costly memory transfer bottleneck.
We demonstrate the effectiveness of the algorithm on a model motivated by next generation sequencing data, and we release our implementation in R packages \fbseq{} and \fbseqCUDA{}.

%% file: body.tex
% !TEX root = ../computation_paper.tex

\section{Introduction} \label{PAPER1sec:introduction}

\begin{figure}[htbp]
   \captionsetup{width=0.8\textwidth} % requires the caption package
   \centering
   \includegraphics[scale=.5]{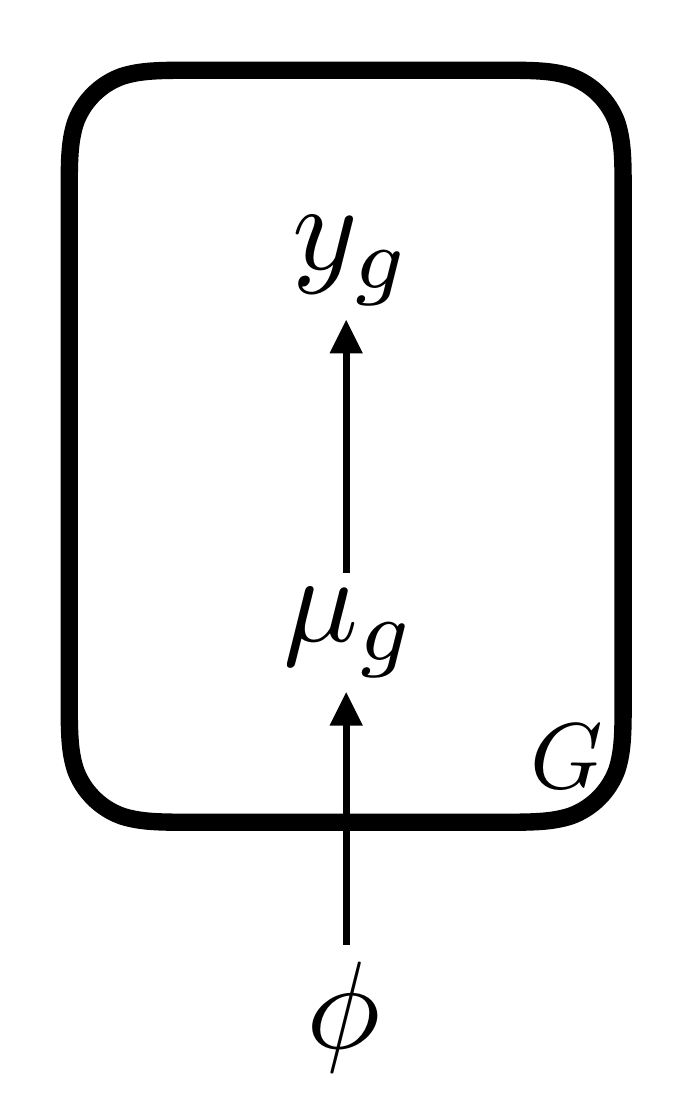} % requires the graphicx package
   \caption{Directed acyclic graph (DAG) representation of a two-level hierarchical model. The box with $G$ in the corner indicates nodes $\mu_g$ and $y_g$ for $g = 1, \ldots, G$ where the node $\phi$ has a directed edge to each $\mu_g$ and each $\mu_g$ has a directed edge to the associated $y_g$.} 
   \label{PAPER1fig:dag1}
\end{figure}

A two-level hierarchical model has the form,
\begin{equation} 
y_g|\mu_g \ind p(y_g|\mu_g),
\qquad \mu_g|\phi \ind p(\mu_g|\phi) \label{PAPER1eq:model}
\end{equation}
where $y_g$ may be a scalar or vector, $y = \{y_1, \ldots, y_G\}$ is the collection observed data, each $\mu_g$ may be a scalar or vector, ${\mu}  = \{ \mu_1 \cdots \mu_G\}$ is the collection of group-specific parameters, $\phi$ is the vector of hyperparameters, and $\ind$ indicates conditional independence. 
Figure \ref{PAPER1fig:dag1} displays a directed acyclic graph (DAG) representation of this model. 
Given a prior $\phi \sim p(\phi)$, our goal is to obtain the full joint posterior density of the parameters, % i.e.\ 
$p(\mu, \phi|y)$. 
Typically, this posterior is analytically intractable, so approximation techniques are used. 
Most commonly, a Markov chain Monte Carlo (MCMC) algorithm such as Metropolis-Hastings, Hamiltonian Monte Carlo, slice sampling, Gibbs sampling, or a combination of these or other techniques are used to obtain samples that converge to draws from this posterior. 
If $G$ is large, implementations of MCMC algorithms that estimate $p(\mu, \phi|y)$ can be slow or even computationally intractable. 

The primary motivating context for our work is in the estimation of parameters in a high-dimensional hierarchical model for data from RNA-sequencing (\RNAseq{}) experiments. 
\RNAseq{} experiments measure the expression levels of tens of thousands of genes in a small collection of samples, and they are used to answer important scientific questions in a multitude of fields, such as biology, agriculture, and medicine \citep{mortazavi2008mapping, paschold, tumor}. 
Each gene is observed a relatively small number of times, and the task is to detect important genes according to application-specific criteria. 
A hierarchical model allows data-based borrowing of information across genes and thereby ameliorates difficulties due to the small sample sizes for each gene.
Unfortunately, estimation of parameters in these models via general purpose Bayesian software or custom-built serial algorithms can be computationally slow or even intractable.

The development of parallelized algorithms for Bayesian analysis is an active area of research enhanced by the wide availability of general purpose graphics processing units (GPUs). 
%For high-dimensional problems, gains in computational efficiency are seen when parallelizing evaluation of the likelihood \citep{suchard2009many, white2014gpu}, parallelizing particle calculations in population-based MCMC and sequential Monte Caro \citep{lee2010utility}
\cite{suchard2009many} utilized GPU-acceleration for likelihood calculations in phylogenetic models.
\cite{lee2010utility} described strategies for parallelizing population-based MCMC and sequential Monte Carlo. 
\cite{suchard2010understanding} outlined a strategy for applying parallelized MCMC to fit mixture models in Bayesian fashion. 
\cite{tibbits2011} proposed and implemented parallel multivariate slice sampling. 
\cite{jacob2011using} built a block independence Metropolis-Hastings algorithm to improve estimators while incurring no additional computational expense due to parallelization.
\cite{murray2014} used hundreds of cores to accelerate an elliptical slice sampling algorithm that approximates the target density with a mixture of normal densities constructed by sharing information across parallel Markov chains. 
\cite{white2014gpu} performed MCMC using GPU-acceleration to calculate likelihoods while modeling terrorist activities.
\cite{gramacy2014} applied multiple high-performance parallel computing paradigms to accelerate Gaussian process regression. 
\cite{beam2015fast} built a GPU-parallelized version of Hamiltonian Monte Carlo in the context of a multinomial regression model.
\cite{gruber2016gpu} utilized GPU-parallelized importance sampling and variational Bayes for estimation and prediction in dynamic models.

We develop a fully Bayesian approach for analyses that use high-dimensional hierarchical models made feasible by the development of efficient parallelized algorithms.
Section \ref{PAPER1sec:parmcmc} develops a strategy for designing parallel MCMC algorithms for estimating full joint posterior distributions of hierarchical models. 
Section \ref{PAPER1sec:gpu} suggests that graphics processing units (GPUs) offer the most appropriate parallel computing platform, and this section explains how to maximize the effectiveness of GPUs. 
Section \ref{PAPER1sec:rnaseq} describes an application of our strategy in the analysis of \RNAseq{} data along with its implementation, a pair of publicly-available R packages. 
Finally, Section \ref{PAPER1sec:runtime} explores the speed of the implementation for both a real dataset and a collection of simulated datasets.

\section{Parallelized MCMC} \label{PAPER1sec:parmcmc}

In most cases, the joint full posterior density $p(\mu, \phi |y)$ for the model in Equation \eqref{PAPER1eq:model} (Figure \ref{PAPER1fig:dag1}) cannot be found analytically, so Markov chain Monte Carlo (MCMC) is often used to obtain samples that converge to draws from this posterior.
A two-step Gibbs sampler involves alternately sampling $\mu$ from its full conditional, $p(\mu|\ldots)$, and $\phi$ from its full conditional, $p(\phi|\ldots)$, where `$\ldots$' indicates all other parameters and the data.
If these full conditionals have no known form, then the Gibbs step is typically replaced with a Metropolis-Hastings, rejection sampling, or slice sampling step \citep{bda, neal2003}. 

For high-dimensional group-specific parameters $\mu_g$ and hyperparameters $\phi$, it is often impractical to sample the entire vector $\mu$ or $\phi$ jointly. 
In these scenarios, the group-specific parameters and hyperparameters are decomposed into subvectors $\mu_g=(\mu_{g1},\ldots,\mu_{gJ})$ and $\phi=(\phi_1,\ldots,\phi_K)$, respectively. 
The component-wise MCMC then proceeds by sampling from these lower-dimensional full conditionals using composition, random scan, or random sequence sampling \citep{johnson2013component}. 

Parallelism increases the efficiency of these MCMC approaches in hierarchical models by simultaneous sampling when parameters are conditionally independent and using parallelized reductions when full conditionals depend on low-dimensional summaries of other parameters.
For hierarchical models, each MCMC step uses conditional independence, reductions, or both, and this designation partitions steps into classes.
When the number of groups $G$ is large, conditional independence can lead to a $G$-fold speedup while parallelized reductions can give a speedup of $\approx G/\log_2(G)$.

\subsection{Simultaneous steps for conditionally independent parameters} \label{PAPER1subsec:simulgibbs}

In the two-level hierarchical model of Equation \eqref{PAPER1eq:model}, the group-specific parameters $\mu_g$ are conditionally independent since 
\[
p(\mu|\ldots) \propto \prod_{g=1}^G p(y_g|\mu_g) p(\mu_g|\phi) \propto \prod_{g=1}^G p(\mu_g|y_g,\phi).
\]
The theory of DAGs also reveals this conditional independence, specifically in nodes that are \emph{d-separated} given the conditioning nodes \citep[Ch. 3]{koller2009probabilistic}.
In Figure \ref{PAPER1fig:dag1}, the nodes $\mu_1,\ldots,\mu_G$ are d-separated given $\phi$ and therefore conditionally independent. % WL: DAGs and full conditionals are 2 different ways to find conditional independence.
Thus the vectors $\mu_g$ can be sampled simultaneously and in parallel.

Often it is more convenient to sample subvectors of the vector $\mu_g$. The $j$th subvector is conditionally independent across $g$ since 
$
p(\mu_{gj}|\ldots) \propto p(y_g|\mu_g)p(\mu_g|\phi).
$
Hence, we sample the $\mu_g$'s, or $\mu_{gj}$'s, in parallel, simultaneous Gibbs steps. 

Parallel execution is accomplished by assigning each group parameter to its own independent unit of execution, or thread. 
With $G$ simultaneous threads, parallelizing across these groups has a theoretical maximum $G$-fold speedup relative to a serial implementation. For \RNAseq{} data analysis with $G \approx 40000$, this is a sizable improvement. 

In addition to conditional independence, the full conditional for a group-specific parameter $\mu_g$ or $\mu_{gj}$ depends only on the data $y_g$ for that group (and the hyperparameters $\phi$). 
Thus, when performing parallel operations, memory transfer is minimized since only a small amount of the total data will need to be accessed by each parallel thread.

Our approach to parallelizing Gibbs steps is a special case of ``embarrassing parallel" computation, which is parallelism without any interaction (i.e.\ data transfer or synchronization) among simultaneous units of execution. Embarrassingly parallel computation is already utilized in existing applications of GPU computing in the acceleration of Bayesian computation. For instance, the strategy by \cite{jacob2011using} shows how embarrassingly parallel computation can accelerate independence Metropolis-Hastings. Much of Metropolis-Hastings is unavoidably sequential because, as with any Monte Carlo algorithm, the value at the current state depends on the value at the previous state. However, in independence Metropolis-Hastings, each proposal draw is generated independently of the previous one, so all proposals can be calculated beforehand in embarrassingly parallel fashion using simultaneous independent threads. Similarly, the Metropolis-Hastings acceptance probability (Figure 1, \cite{jacob2011using}) at each step contains a factor that depends only on the current proposal, and these factors can similarly be computed in parallel. 

Parallelization of importance sampling is similarly straightforward, as in Figure 2 of \cite{lee2010utility}. The strategy by \cite{gruber2016gpu}, which uses a decoupling/recoupling strategy to fit dynamic linear models of multivariate time series, takes advantage of embarrassingly parallel computation both within and among importance samplers. At each time point, they parallelize across the non-temporal dimension to draw Monte Carlo samples separately from independent prior distributions in their model (Section 3-B), and then parallelize across both the non-temporal dimension and the Monte Carlo sample size to draw from approximate posterior distributions (Section 3-C,D).

\subsection{Reductions to aid the efficiency of hyperparameter sampling} \label{PAPER1subsec:redhyper}

The hyperparameter full conditionals, $p(\phi|\ldots)$ or $p(\phi_k|\ldots)$, usually depend on \emph{sufficient quantities} that act as sufficient statistics of $\mu$. 
For example, if $\mu_g\ind N(\phi_1,\phi_2^2)$, then the sum of $\mu_g$ and the sum of $\mu_g^2$ over index $g$ are minimal sufficient for $\phi = \left (\phi_1, \phi_2 \right )$.
More generally, if $p(\mu_g|\phi)$ is an exponential family (or generalized linear model), then there is a sufficient quantity that depends on the model matrix (design matrix) and $\mu$ \cite[Ch. 2]{mccullagh1989generalized}. % [pg. 32]

Each sufficient quantity %, i.e.\ sum or product of components of $\mu$,
can be computed using a \emph{reduction}, i.e.\ repeated application of a binary operator to pairs of $\mu_g$'s until a single scalar is returned.
A serial application of a reduction over $G$ quantities requires $G - 1$ operations. 
In contrast, a parallelized reduction over $G/2$ threads has complexity $\log_2(G)$. 
For large $G$, the speedup is considerable, so parallelizing the reductions on $\mu$ speeds up the sampling of the hyperparameter full conditionals. 
For example, for \RNAseq{} data analysis with $G \approx 40000$, a parallelized reduction provides a theoretical speedup of $\frac{G - 1}{\log_2(G)} \approx 2600$. Of course, the observed efficiency gain depends on the software implementation, and many parallel computing frameworks have built-in optimized reduction functionality. CUDA's Thrust library, for example, allows the user to perform a fast parallelized reduction with a single line of code \citep{thrust}.

Steps requiring reductions can be identified in a DAG where nodes have directed edges outward. % WL: I tried to make this clearer, and I hope I preserved the intended meaning.
When the number of edges from a node is large, a parallelized reduction is beneficial.
For the two-level hierarchical model of Figure \ref{PAPER1fig:dag1}, the node $\phi$ has $G$ exiting edges, and when $G$ is large, the corresponding sampler benefits from a parallelized reduction.
In the case where $\mu_g$ is of low-dimension and $y_g$ is relatively large, e.g.\ $y_{gj}\ind N(\mu_g,\sigma^2)$ for $j=1,\ldots,J$ where $J$ is large, a parallelized reduction for each $\mu_g$ may also be beneficial.

Reductions are used in other GPU-accelerated Bayesian analyses and Markov chain Monte Carlo routines. As an example, consider the GPU-accelerated Gaussian process modeling method by \cite{gramacy2014}. A major goal is to generate a large set of predictions, where each prediction is computed using a different subset of the available data. Each of these optimal subsets is determined with a criterion equivalent to mean squared prediction error, and the computation of this criterion, which depends on quadratic forms involving the correlation matrix, is expensive. As part of the acceleration, \cite{gramacy2014} use parallelized pairwise summation in the calculation of these quadratic forms. Rather than Thrust, their implementation uses the parallelized reduction method by \cite{sharcnet2012}. For other examples of parallelized reductions in Bayesian methods and MCMC, see \cite{suchard2010understanding} and \cite{suchard2009many}.

\subsection{More hierarchical levels} \label{PAPER1sec:threelevel}

\begin{figure}[htbp]
   \captionsetup{width=0.8\textwidth} % requires the caption package
   \centering
   \includegraphics[scale=.5]{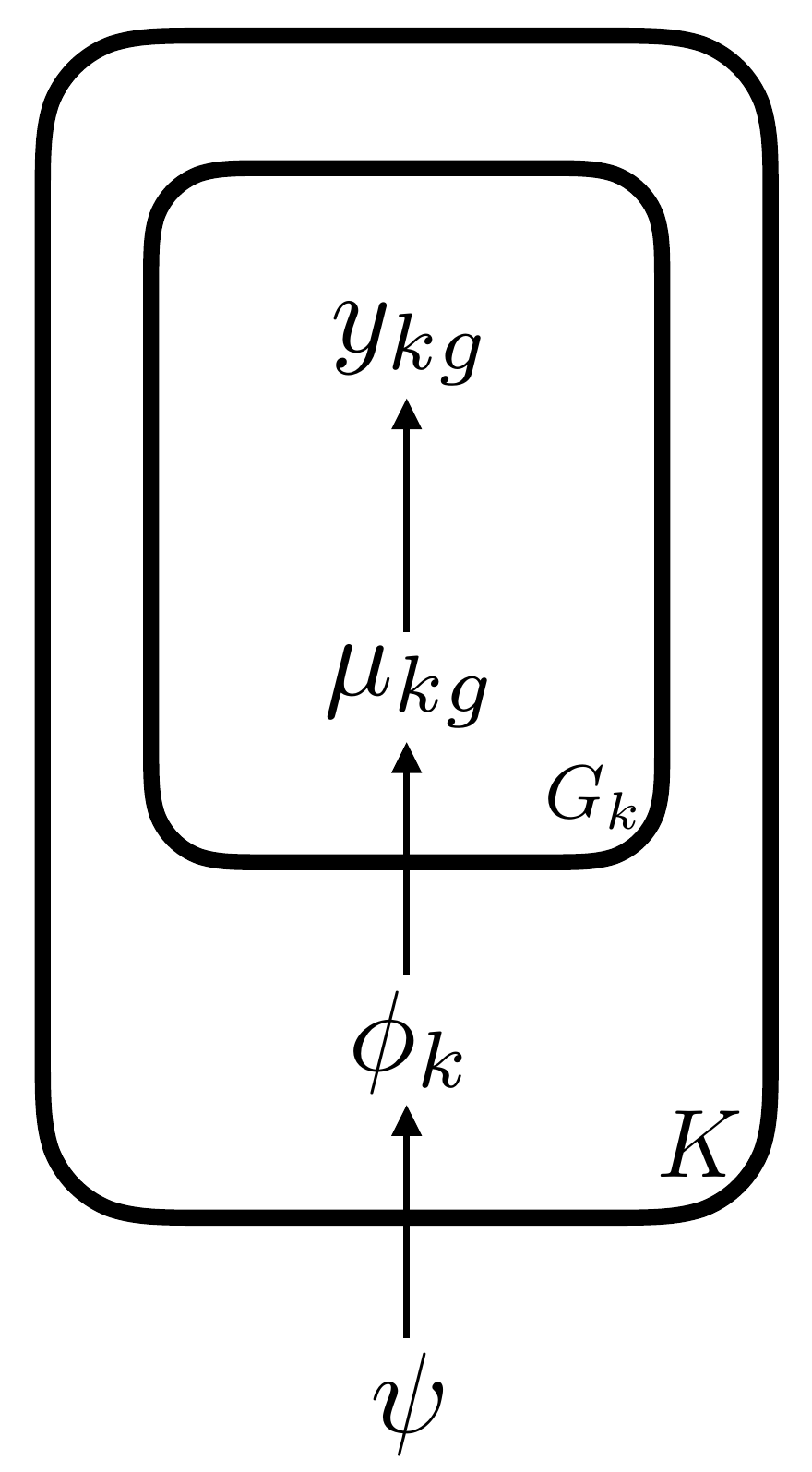} % requires the graphicx package
   \caption{Directed acyclic graph (DAG) representation of the three-level hierarchical model in Section \ref{PAPER1sec:threelevel}. The box with $K$ indicates replication over $k$, and the box with $G_k$ indicates replication over $g$.} 
   \label{PAPER1fig:dag2}
\end{figure}

Dichotomizing Gibbs steps into those that benefit from conditional independence and those that benefit from parallelized reductions extends to additional levels of hierarchy.
Consider the three-level hierarchical model 
\begin{equation} 
y_{kg} \ | \ \mu_{kg} \ind p(y_{kg}|\mu_{kg}), \quad 
\mu_{kg} \ | \ \phi_k \ind p(\mu_{kg}|\phi_k), \quad \mbox{and}\quad 
\phi_{k} \ | \ \psi \ind p(\phi_k|\psi) % WL: \phi => \phi_k
\label{PAPER1eq:model2}
\end{equation}
where $k=1,\ldots, K$, $g=1,\ldots,G_k$, and $y_{kg}$, $\mu_{kg}$, $\phi_k$, and $\psi$ could all be vectors. Figure \ref{PAPER1fig:dag2} displays a DAG representation of the model in Equation \eqref{PAPER1eq:model2}.
A two-step Gibbs sampler for this model alternately samples 
\[ 
\mu,\psi \sim p(\mu|y,\phi)p(\psi|\phi) \quad \mbox{and} \quad 
\phi \sim p(\phi|y,\mu,\psi)
\]
which shows that $\mu$ and $\psi$ are conditionally independent given $\phi$.
The components of $\mu$ are conditionally independent, as well as the components of $\phi$, since 
\[ 
p(\mu|\ldots) \propto \prod_{k=1}^K \prod_{g=1}^{G_k} p(y_{kg}|\mu_{kg}) p(\mu_{kg}|\phi_k) \quad \mbox{and} \quad
p(\phi|\ldots) \propto \prod_{k=1}^K \prod_{g=1}^{G_k} p(\mu_{kg}|\phi_k) p(\phi_k|\psi).
\]
Figure \ref{PAPER1fig:dag2} also displays these conditional independencies: $\psi$ and $\mu_{kg}$ ($k = 1, \ldots, K$, $g = 1, \ldots, G_k$) are d-separated given $y$ and $\phi$, and the $\phi_k$'s are d-separated given $\mu$ and $\psi$. 

As before, the full conditional of $\phi_k$ depends on a sufficient quantity calculated from $\{\mu_{k1},\ldots,\mu_{kG_k}\}$ and the full conditional of $\psi$ depends on a sufficient quantity calculated from $\phi$. 
Figure \ref{PAPER1fig:dag2} displays this relationship as well since there are 1) many edges from $\psi$ to the $\phi_k$'s, and 2) many edges from $\phi_k$ to the $\mu_{kg}$'s. 

If $K$ and $G_k$ for $k=1,\ldots,K$ are large, then parallelizing these conditional independencies and calculations of sufficient quantities will dramatically improve computational efficiency. When additional levels are added to the hierarchy, each full conditional can be categorized into a conditional independence step, a parallelized reduction step, or both.

\section{Acceleration with high-performance computing} \label{PAPER1sec:gpu}

Three general parallel computing architectures currently exist: multi-core/CPU machines, clusters, and accelerators. Modern computers have multiple CPUs, each with multiple cores, where each core can support one or more parallel threads or processes. This multi-core/CPU hardware allows fast communication among threads, as the threads have abundant shared memory. However, this paradigm only supports tens of simultaneous threads, not hundreds or thousands, and thus will not provide the desired efficiency gain. In contrast, clusters, or collections of networked computers, provide the possibility of unlimited parallelism, but communication of threads occurs across a relatively slow network. Between these extremes lie accelerators such as NVIDIA CUDA graphics processing units (GPUs) and Intel MIC coprocessors. GPUs in particular are capable of spawning hundreds of thousands of threads at a time, and these threads are partitioned into groups called blocks \citep{cuda}. Each block can contain hundreds of threads, and communication among the threads in a single block is extremely fast, driving the acceleration of reductions even when several blocks are needed. 
%CUDA's Thrust library makes parallelized reductions easy \citep{thrust}. 

% To implement parallelize MCMC, it is best to choose a technology that allows for the use of thousands to tens of thousands of simultaneous units of execution, or threads, where the computational load on each individual thread is relatively light and large sub-collections of threads can share information quickly. Traditional CPU multi-core computing allows threads to communicate with each other fast, but the number of threads is extremely small. Cluster computing across multiple computers or nodes can support a massive number of simultaneous threads, but without GPUs, there are few threads per node, and threads in different nodes communicate slowly. In general-purpose GPU computing, however, threads are both abundant and interconnected. A graphics processing unit, or GPU, has thousands of computing cores, and is capable of spawning hundreds of thousands of threads at a given time . 

However, if GPUs are used, the implementation strategy needs to be optimized for GPU computing. In particular, it is important to minimize the amount of data transferred between CPU memory and GPU memory \citep{beam2015fast}. In our applications, this data transfer is by far the most time-consuming step, and misuse can easily defeat the purpose of GPU computing altogether. In particular, copying all MCMC parameter samples from GPU memory to CPU memory would be intractably slow. In addition, it is important to avoid exhausting all available GPU memory. There are opportunities to make GPU computing effective throughout the whole analysis.

\subsection{Cumulative means and means of squares} \label{PAPER1subsec:cummean}

Using cumulative means on the GPU, keep track of the mean and mean square of each parameter's MCMC samples, separately for each MCMC chain in the analysis. More specifically, suppose $C$ independent MCMC chains with $M$ iterations each are used to estimate some joint posterior distribution. In addition, let $\theta$ be an arbitrary parameter and $\theta_c^{(m)}$ be the the $m$'th MCMC sample of $\theta$ in chain $c$, where $m = 1, \ldots, M$ and $c = 1, \ldots, C$. Using a one-pass algorithm \citep{ling} over the course of the MCMC, record each $\ov{\theta}_c= \frac{1}{M}\sum_{m = 1}^M \theta_c^{(m)}$ and $\ov{\theta^2}_c = \frac{1}{M} \sum_{m = 1}^M \left(\theta_c^{(m)} \right)^2$. One option for the computation of $\ov{\theta}_c$ is to update the cumulative sum $\sum_{i = 1}^m \theta_c^{(i)}$ on each MCMC iteration $m$ and divide by $M$ at the end, an approach that may suffer a loss of precision in some applications. A one-pass algorithm due to \cite{welford}, on the other hand, which updates $x_{m - 1} = \frac{1}{m - 1} \sum_{i = 1}^{m - 1} \theta_c^{(i)}$ to $x_m = x_{m - 1} + \frac{\theta_c^{(m)} - x_{m - 1}}{m}$ on iteration $m$, is more numerically stable. 
%For more information on algorithms for computing means and variances, see \cite{ling}.

The quantities $\ov{\theta}_c$ and $\ov{\theta^2}_c$ ($c = 1, \ldots, C$) have two major uses. The first is for assessing convergence via the Gelman-Rubin potential scale reduction factor \citep{bda} for $\theta$, which is given by
\[
\widehat{R} = \sqrt{1 + \frac{1}{M} \left(\frac{B}{W} - 1 \right)}
\] 
where 
\[
B = \frac{M}{C - 1} \sum_{c = 1}^C  \left (\ov{\theta}_c - \ov{\theta} \right )^2,
W = \frac{1}{C} \sum_{c = 1}^C S_c^2,
\]
\[
\ov{\theta} = \frac{1}{C} \sum_{c = 1}^C \ov{\theta}_c,
\mbox{ and }
S_c^2 = \frac{M}{M-1}\left[\ov{\theta^2}_c-\ov{\theta}^2_c \right] \approx \ov{\theta^2}_c-\ov{\theta}^2_c 
%\frac{1}{M - 1} \sum_{m = 1}^M \left (\theta_c^{(m)} - \ov{\theta}_c \right )^2
.
\]
A Gelman factor $\widehat{R}$ far above 1 is evidence of lack of convergence in $\theta$. It is a recommended and common practice to run at least 4 MCMC chains, starting at parameter values overdispersed relative to the full joint posterior distribution, and then check that the Gelman factors of parameters of interest are below 1.1 before moving forward with the analysis.
The use of cumulative means allows the calculation of Gelman factors, and therefore convergence assessment, without the need to return all parameter samples.

%\subsection{Parameter estimation}

The second main use of the cumulative mean and mean of squares is for point and interval estimates.
%Point and interval estimates for $\theta$ can be calculated from $\ov{\theta}$ and $\ov{\theta^2} = \frac{1}{C} \sum_{c = 1}^C \ov{\theta^2}_c$. 
By the Strong Law of Large Numbers, $\ov{\theta}$ and $\ov{\theta^2}$ converge almost surely to the expected values $E(\theta|y)$ and $E(\theta^2|y)$, respectively. Thus, $\ov{\theta}$ and $\ov{\theta^2}-\ov{\theta}^2$ are MCMC approximations to the posterior mean and variance, respectively. 
Since the posterior distribution itself converges to a normal distribution for large amounts of data (the primary use case for the computational methods developed here), a $100(1-\alpha)$\% approximate equal-tail credible interval can be constructed via $\ov{\theta} \pm z_{\alpha/2} \sqrt{\ov{\theta^2}-\ov{\theta}^2}$ where $P(Z>z_\alpha)=\alpha$ and $Z$ is a standard normal distribution.

% For example, if we can safely assume that a normal distribution well approximates marginal posterior $p(\theta|y)$, then an approximate 95\%-credible for $\theta$ is
% 
% $(z_\ell, z_U)$, where
% \begin{align*}
% \int_{-\infty}^{z_\ell} f(\theta) d \theta = 0.025 \qquad \int_{-\infty}^{z_U} f(\theta) d \theta = 0.975
% \end{align*}
% and 
% \begin{align*}
% f(x) = \frac{1}{\sqrt{2 \pi \sigma^2}} \exp \left (\frac{1}{2\sigma^2} \left (x - \mu \right )^2 \right ) \qquad \mu = \ov{\theta} \qquad \sigma^2 = \frac{M}{M-1} \left (\ov{\theta^2} - \ov{\theta}^2 \right )
% \end{align*}

With approximate credible intervals for each model parameter, it is typically not necessary to save all MCMC parameter samples.
To reduce data transfer between the GPU and CPU, we recommend copying only a select few parameter samples back to CPU memory for future use, preferably the hyperparameters $\phi$ and some group-specific parameter samples $\mu_{gj}$ for a select, perhaps random, few values of $g$ and $j$. For those parameters, an appropriate thinning interval should be used, although the cumulative means and means of squares should be calculated using all samples, and the GPU should retain only a single iteration at any given time. That way, memory-based computational bottlenecks are avoided, and enough parameter samples will be available post-hoc for checking the distributional assumptions of the approximate credible intervals.

\subsection{Inference}

Similar to the point and interval estimates in Section \ref{PAPER1subsec:cummean}, inferential quantities that depend on $\mu$ should be calculated using cumulative means instead of the full collection of MCMC parameter samples. For example, a posterior probability that can be expressed as $P(f(\mu, \phi) \ | \ y)$ should be estimated by $\frac{1}{M} \sum_{m = 1}^M \I\left ( f\left(\mu^{(m)}, \phi^{(m)} \right ) \right )$, where $f$ is any function of the parameters that returns a true/false value, $\I(\cdot)$ is the indicator function, and the mean of indicator functions can be calculated using a one-pass algorithm as in Section \ref{PAPER1subsec:cummean}. 

Unfortunately, most parallel computing tools operate at a low level, so it is generally impossible to allow the user to specify a generic function $f$. However, posterior probabilities involving contrasts are straightforward to implement. Such a probability is of the form, $P \left ( u_1^T \eta > b_1 \text{ and } \ldots \text{ and } u_K^T \eta > b_K \ | \ y \right )$, where $\eta$ is the vector obtained by concatenating the $\mu_g$ vectors and $\phi$, each fixed vector $u_k$ ($k = 1, \ldots, K)$ has the same length as $\eta$, and $b_1, \ldots, b_K$ are fixed scalars.\footnote{
Practitioners may desire $P \left ( u_1^T \eta > b_1 \otimes_1 \ldots \otimes_K u_K^T \eta > b_K \ | \ y \right )$, where each $\otimes_k$ could be either ``and" or ``or". This general form can be obtained from probabilities using only ``and" along with the general disjunction rule in basic probability theory. For example, $P \left ( u_1^T \eta > b_1 \text{ or } u_1^T \eta > b_1 \right ) = 
P \left ( u_1^T \eta > b_1 \right ) + P \left ( u_2^T \eta > b_2 \right ) - P \left ( u_1^T \eta > b_1 \text{ and } u_1^T \eta > b_1 \right )$. The probabilities on the right are estimated using a one-pass algorithm during the MCMC, and then the estimate on the left is calculated afterwards. This restriction to ``and" in the main program simplifies the implementation.
} 
The MCMC estimate is \\ $\frac{1}{M} \sum_{m = 1}^M \I \left ( u_1^T \eta^{(m)} > b_1 \text{ and } \ldots \text{ and } u_K^T \eta^{(m)} > b_K \right )$, where $\eta^{(m)}$ is the MCMC sample of $\eta$ at iteration $m$. For probabilities specific to each $\mu_g$, if the $\mu_g$'s are all of the same length, the formulation is \\ $P \left ( v_1^T \mu_g > b_1 \text{ and } \ldots \text{ and } v_K^T \mu_g > b_K \ | \ y \right )$ for $g = 1, \ldots, G$, where $v_1, \ldots, v_K$ are fixed vectors of the same length as $\mu_1$. In this case, the estimate for index $g$ is  \\ $ \frac{1}{M} \sum_{m = 1}^M \I \left (v_1^T \mu_g^{(m)} > b_1 \text{ and } \ldots \text{ and } v_K^T \mu_g^{(m)} > b_K \right)$, and these estimates can be updated in parallel over index $g$. An example of this last construction is \\ $P \left ( \mu_{g2} + \mu_{g4}> 0 \text{ and } \mu_{g3} + \mu_{g4}> 0 \ | \ y \right )$ for $g = 1, \ldots, G$, estimated by  \\
$\frac{1}{M} \sum_{m = 1}^M \I \left(\mu_{g2}^{(m)} + \mu_{g4}^{(m)}>0 \text{ and } \mu_{g3}^{(m)} + \mu_{g4}^{(m)}> 0 \right )$ for a given $g$. These posterior probabilities often arise in \RNAseq{} data analysis where the goal is often to detect genes with important patterns in their expression levels.

\section{Application to RNA-sequencing data analysis} \label{PAPER1sec:rnaseq}

We apply the above strategy to RNA-sequencing (\RNAseq{}) data analysis. \RNAseq{} is a class of next-generation genomic experiments that measure the expression levels of genes in organisms across multiple groups or experimental conditions. The data from such an experiment is a matrix of counts, where the count in row $g$ and column $n$ is the relative expression level of gene $g$ found in \RNAseq{} sample $n$. For a more detailed, technical description of \RNAseq{} experiments and data preprocessing, see \cite{datta2014statistical}, \cite{oshlack2010}, and \cite{wang2010}.

The goal of the analysis is to model gene expression levels and detect important genes, a difficult task because there are typically $G \approx 40000$ genes and $N \approx 10$ \RNAseq{} samples. Hierarchical models are suitable because they borrow information across genes to improve detection. However, fitting them is computationally demanding because of the high number of genes and low number of observations per gene. Many approaches ease the computation with empirical Bayes methods, where the hyperparameters $\phi$ are set constant at values calculated from the data that approximate the respective target densities before the MCMC begins (\cite{hardcastle2012bayseq}; \cite{ji2014estimation}; \cite{niemi2015}). However, empirical Bayes approaches ignore uncertainty in the hyperparameters, so a fully Bayesian solution may be preferred.

\subsection{Model} \label{PAPER1subsec:model}

%In this section, we give a detailed explanation of our model. For a compact summary, along with a directed acyclic graph (DAG) representation, please see Figure \ref{PAPER1fig:dag3}.

Let $y_{g n}$ be the \RNAseq{} count for sample $n$ ($n = 1, \ldots, N)$ and gene $g$ ($g = 1, \ldots, G)$.
Let $X$ be the $N \times L$ model matrix for gene-specific effects ${\beta}_g = (\beta_{g 1}, \ldots, \beta_{g L})$. 
Let ${X}_n$ be the $n^{th}$ row of $X$. 
We assume 
$y_{g n} | \varepsilon_{g n}, {\beta}_g \ind \text{Poisson} \left (\exp \left (h_n + \varepsilon_{g n} + {X}_n {\beta}_g \right ) \right )$. The $h_n$'s are constants estimated from the data, and they take into account sample-specific nuisance effects such as sequencing depth (\cite{amap}, \cite{anders2010differential}, \cite{robinson2010scaling}). The $\varepsilon_{g n}$ parameters account for overdispersion, and we assign $\varepsilon_{g n} | \gamma_g \ind \text{Normal}(0, \gamma_g)$.
The $\gamma_g$ parameters are analogous to the typical gene-specific negative-binomial dispersion parameters used in many other methods of \RNAseq{} data analysis \citep{landau2013dispersion}. 
We assign $\gamma_g | \tau, \nu\ind \text{Inverse-Gamma} \left (\nu/2, \nu\tau/2 \right )$. $\tau$ is a prior measure of center of the $\gamma_g$ terms (between the prior mean and the prior mode), and $\nu$ is the degree to which the $\gamma_g$'s ``shrink" towards $\tau$. We assign $\tau \sim \text{Gamma}(a, \text{rate} = b)$ and $\nu \sim \text{Uniform}(0, d)$, where $a = 1$, $b = 1$, and $d = 1000$ are fixed constants such that these priors are diffuse \citep{gelman2006prior}. % WL: values of constants a, b, d in priors

\begin{figure}[htbp]
   \centering
   \begin{minipage}{0.49\textwidth}
   \begin{align*}
&y_{gn} \stackrel{\text{ind}}{\sim} \text{Poisson} \left (\exp \left (h_n + \varepsilon_{gn} + X_n \beta_{g} \right ) \right ) \\
& \qquad \varepsilon_{gn} \stackrel{\text{ind}}{\sim} \text{Normal}(0, \gamma_g) \\
& \qquad \qquad \gamma_g \stackrel{\text{ind}}{\sim} \text{Inverse-Gamma}\left (\frac{\nu}{2}, \frac{\nu \tau}{2} \right) \\
& \qquad \qquad \qquad \nu \stackrel{\text{}}{\sim} \text{Uniform}(0, d) \\
& \qquad \qquad \qquad \tau \stackrel{\text{}}{\sim} \text{Gamma}(a, \text{rate} = b) \\
& \qquad \beta_{g\ell} \stackrel{\text{ind}}{\sim} \text{Normal}(\theta_\ell, \sigma_\ell^2) \\
& \qquad \qquad \theta_\ell \stackrel{\text{ind}}{\sim} \text{Normal}(0, c_\ell^2) \\
& \qquad \qquad \sigma_\ell \stackrel{\text{ind}}{\sim} \text{Uniform}(0, s_\ell)
\end{align*}
   \end{minipage} 
   \begin{minipage}{0.49\textwidth}
   \includegraphics[scale=.5]{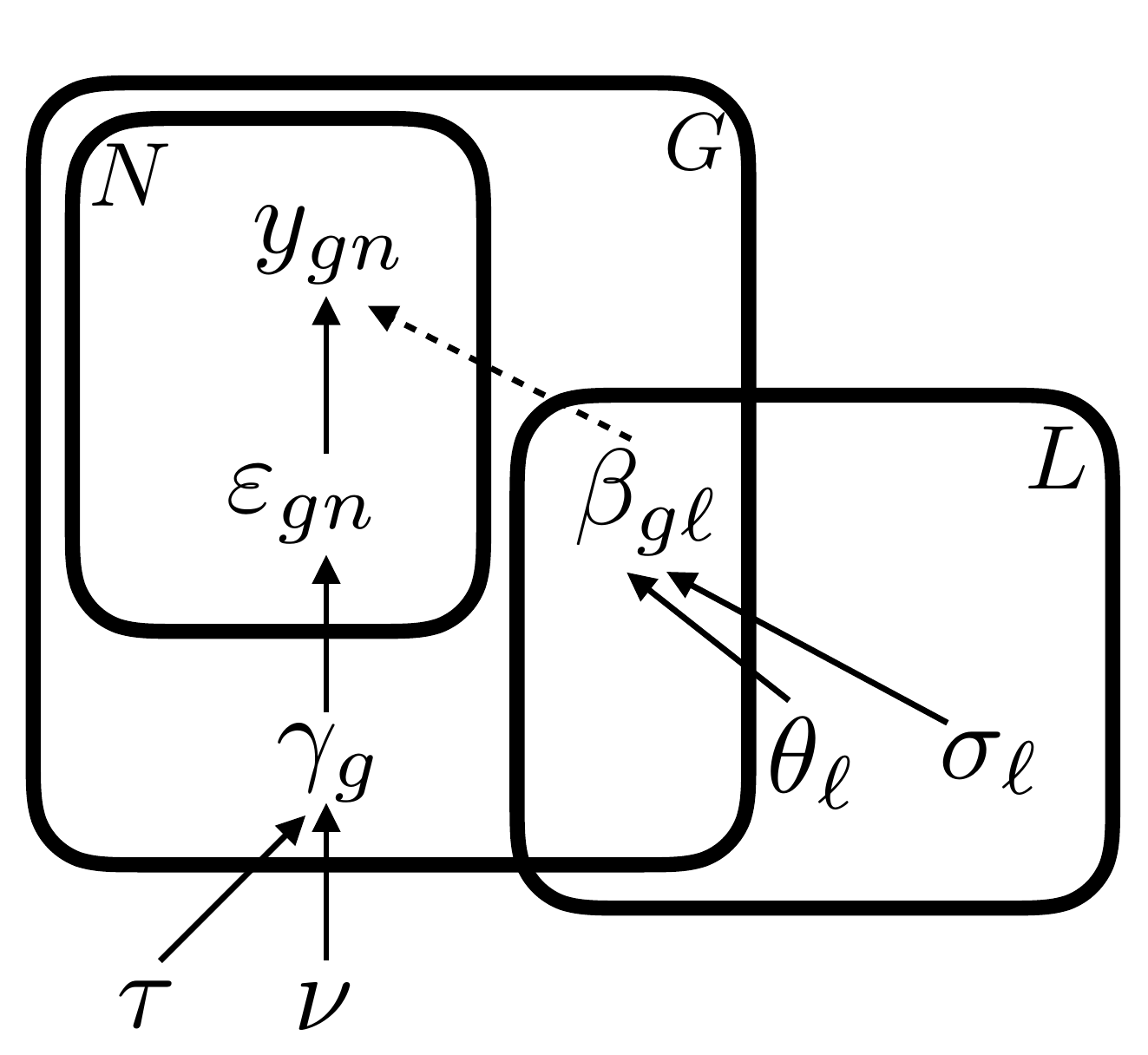}
   \end{minipage}
   \caption{Directed acyclic graph (DAG) representation of the \RNAseq{} model in Section \ref{PAPER1subsec:model}, along with a formulaic representation on the left. The box with $G$ in the corner indicates that each parameter inside represents multiple nodes, each specific to a value of $g = 1, \ldots, G$. The analogous interpretation holds for the boxes with $N$ and $L$, respectively. The dashed arrow from $\beta_{g\ell}$ to $y_{gn}$ indicates that an edge is present if and only if $X_n \beta_g$ is a non-constant function of $\beta_{g \ell}$: that is, if and only if $X_{n\ell} \ne 0$, where $X$ is the model matrix and $X_n$ is its $n$'th row.} 
   \label{PAPER1fig:dag3}
\end{figure}

The ${\beta}_g$ terms relate elements of the model parameterization to gene expression levels, and we interpret ${X}_n {\beta}_g$ to be the log-scale mean expression level of gene $g$ in \RNAseq{} sample $n$. For each fixed $\ell$ from 1 to $L$, we assign $\beta_{g \ell} | \theta_\ell, \sigma_\ell \ind \text{Normal}(\theta_\ell,  \sigma_\ell^2)$. Lastly, we assign $\theta_\ell \ind \text{Normal}(0, c_\ell^2)$ and $\sigma_\ell \ind \text{Uniform}(0, s_\ell)$, where $c_\ell = 10$ and $s_\ell = 100$ ($\ell = 1, \ldots, L$) are fixed constants so that these priors are diffuse \citep{gelman2006prior}. % WL: values of constants c_\ell and s_\ell in priors
This model is summarized and depicted as a DAG in Figure \ref{PAPER1fig:dag3}.

The conditional independence of the $\beta_{g \ell}$'s depends on the model matrix $X$. Parameters $\beta_{1\ell}, \ldots, \beta_{G \ell}$ are always conditionally independent given $\theta_\ell$ and $\sigma_\ell$, but $\beta_{g i}$ and $\beta_{g j}$ are not necessarily conditionally independent for $i \ne j$. To see this, it is easiest to refer to the directed acyclic graph (DAG) representation of the model in Figure \ref{PAPER1fig:dag3}. The dashed arrow from $\beta_{g\ell}$ to $y_{gn}$ indicates that an edge is present if and only if $X_n \beta_g$ is a non-constant function of $\beta_{g \ell}$: that is, if and only if $X_{n\ell} \ne 0$. If there exists any integer $n$ from 1 to $N$ such that there is a directed edge from $\beta_{g i}$ to $y_{gn}$ and another directed edge from $\beta_{g j}$ to $y_{gn}$, then $\beta_{g i}$ and $\beta_{g j}$ are not conditionally independent: here, $y_{gn}$ is a collider on an undirected path between $\beta_{g i}$ and $\beta_{g j}$, making $\beta_{g i}$ and $\beta_{g j}$ not d-separated in the DAG given the other nodes. If no such $n$ exists, then $\beta_{g i}$ and $\beta_{g j}$ are d-separated given the other nodes and thus conditionally independent.

%To elaborate, the full conditional density of $\beta_{g \ell}$ is proportional to
%\begin{align*}
%\exp \left (\beta_{g\ell} \sum_{n = 1}^N y_{gn} X_{n\ell} -  \frac{(\beta_{g\ell} - \theta_\ell)^2}{2 \sigma_\ell^2}  - \sum_{n = 1}^N %\exp \left ( X_{n\ell} \beta_{g\ell} \right ) \exp  \left (h_n + \varepsilon_{gn} +  \sum_{i \ne \ell} X_{ni} \beta_{gi} \right )  \right ) \\
%\end{align*}
%which depends on $\sum_{i \ne \ell} X_{ni} \beta_{gi}$. 

This \RNAseq{} model is a special case of the model in equation \eqref{PAPER1eq:model} and Figure \ref{PAPER1fig:dag1}. To make the transition, note that $(y_{g 1}, \ldots, y_{g N})$ becomes $y_g$, $(\varepsilon_{g 1}, \ldots, \varepsilon_{g N}, \gamma_g,  \beta_{g 1}, \ldots, \beta_{g L})$ becomes $\mu_g$, and $(\tau, \nu, \theta_1, \ldots, \theta_L, \sigma_1, \ldots, \sigma_L)$ becomes $\phi$.

\subsection{MCMC} \label{PAPER1subsec:mcmc}

To fit the model to \RNAseq{} data, we use an overall Gibbs sampling structure and apply the univariate stepping-out slice sampler in Appendix \ref{PAPER1appendix:slice} within each of several Gibbs steps. This versatile slice-sampling-within-Gibbs approach was suggested by \cite{neal2003} (Section 4: Single-variable slice sampling methods), then detailed by \cite{slicewithingibbs} and alluded to by \cite{bda} (Ch 12.3) and \cite{banerjee}. In each of the steps of Algorithm \ref{PAPER1alg:mcmc}, a slice sampler is used to sample from all non-normal full conditionals. Each slice-sampled parameter ($\gamma_1$, $\gamma_2, \varepsilon_{50, 5}$, etc.) has its own tuning variable $w$ and auxiliary variable $w_{\text{aux}}$.

Slice sampling is used for the gamma and inverse-gamma full conditionals in addition to the full conditionals with unknown distributional form. This is because CURAND, the random number generation library for CUDA, has no gamma sampler. Although a gamma sampler could have been implemented (see Appendix B.3 of \cite{gruber2016gpu}), this slice-sampling approach is more versatile. 

\begin{algorithm}
\caption{MCMC for hierarchical \RNAseq{} model}
\label{PAPER1alg:mcmc}
%\begin{algorithmic}
\begin{enumerate}
\item \prl{} sample the $\varepsilon_{g n}$'s.
\item \prl{} sample the $\gamma_g$'s.
\item \red{} to calculate $\sum_{g = 1}^G \left [ \log \gamma_g + \frac{\nu}{\gamma_g} \right ]$. Then sample $\nu$ from its full conditional density, which is proportional to \\
\[
\exp \left (- G \log \Gamma \left (\frac{\nu}{2} \right ) + \frac{G \nu}{2} \log \left ( \frac{\nu \tau}{2} \right )  - \frac{\nu}{2} \sum_{g = 1}^G \left [ \log \gamma_g + \frac{\nu}{\gamma_g} \right ] \right ).
\]
\item \red{} to calculate $\sum_{g = 1}^G \frac{1}{\gamma_g}$. Then sample \\ $\tau\sim \text{Gamma}\left(a + \frac{G \nu}{2}, \ \text{rate} = b + \frac{\nu}{2} \sum_{g = 1}^G \frac{1}{\gamma_g} \right )$.
\item \label{PAPER1itm:beta} For $\ell = 1, \ldots, L$, {\bf in parallel,} sample $\beta_{1\ell}, \ldots, \beta_{G \ell}$. 
\item \label{PAPER1itm:theta} \red{} to calculate means and variances of the relevant $\beta_{g\ell}$'s. Then sample $\theta_1, \ldots, \theta_L$.
\item \label{PAPER1itm:sigma} \red{} to calculate the shape and scale parameters of the inverse-gamma distributions. Then sample $\sigma_1, \ldots, \sigma_L$.
\end{enumerate}
%\end{algorithmic}
\end{algorithm}

In Algorithm \ref{PAPER1alg:mcmc}, we highlight the two types of steps: {\bf in parallel} for the steps with conditionally independent parameters and {\bf reduction} for the parameters whose full conditionals depend on sufficient quantities calculated from other parameters. 
In step \ref{PAPER1itm:beta}, the $\beta_{g\ell}$'s are conditionally independent across $g$ for a given $\ell$, but not necessarily conditionally independent across $\ell$, as the conditional independence of the $\beta_{g\ell}$'s depends on the model matrix.
In steps \ref{PAPER1itm:theta} and \ref{PAPER1itm:sigma}, parameter sampling after the parallelized reductions could be parallelized, but the efficiency gain is small if $L$ is small. In our application, $L$ is $5$. % WL: symbolic cross-referencing of \item entries

\subsection{Implementation}

We release the implementation of this algorithm in R packages \fbseq{} and \fbseqCUDA{}, publicly available on GitHub in repositories named \fbseq{} and \fbseqCUDA{}, respectively. We use two packages for the same method in order to separate the GPU-dependent backend from the platform-independent user interface. \fbseq{} is the pure-R user interface, which is for planning computation and analyzing results on any machine, such as a local office computer. \fbseqCUDA{} is the CUDA-accelerated backend that runs Algorithm \ref{PAPER1alg:mcmc}. The \fbseqCUDA{} package uses custom CUDA kernels (functions encoding parallel execution on the GPU) to run sets of parallel Gibbs steps and CUDA's Thrust library for parallelized reductions. Users can install it on a computing cluster, a G2 instance on Amazon Web Services, or another (likely remote) CUDA-capable resource, and run the algorithm with a function in \fbseq{} that calls the \fbseqCUDA{} engine. For step-by-step user guides, please refer to the package vignettes. We also release \fbseqComputation{}, an R package that replicates the results of this paper. The \fbseqComputation{} package is publicly available through the GitHub repository of the same name. Install \fbseqComputation{} according to the instructions in the package vignette, and run the {\tt paper\_computation()} function to reproduce the computation in Section \ref{PAPER1sec:runtime}.

\section{Assessing computational tractability} \label{PAPER1sec:runtime}

As an example of \RNAseq{} data, we consider the dataset from \cite{paschold}. The underlying \RNAseq{} experiment focused on $N = 16$ biological replicates (pooled from the harvested primary roots of 3.5-day-old seedlings), each from one of 4 genetic varieties, and reported the expression levels of $G = 39656$ genes. The genetic varieties are B73 (an inbred population of Iowa corn), Mo17 (an inbred population of Missouri corn), B73$\times$Mo17 (a hybrid population created by pollinating B73 with Mo17), and Mo17$\times$B73 (a hybrid population created by pollinating Mo17 with B73).

The $N = 16$ by $L = 5$ model matrix is compactly represented as
\begin{align*}
X = \left (
\begin{bmatrix}
1 & \phantom{-}1 & -1 & \phantom{-}0 \\
1 & -1 & \phantom{-}1 & \phantom{-}0 \\
1 & \phantom{-}1 & \phantom{-}1 & \phantom{-}1 \\
1 & \phantom{-}1 & \phantom{-}1 & -1 \\
\end{bmatrix} \otimes
\begin{bmatrix}
1  \\
1  \\
1 \\
1  \\
\end{bmatrix} \qquad
\begin{bmatrix}
1  \\
1  \\
1 \\
1  \\
\end{bmatrix} \otimes
\begin{bmatrix}
\phantom{-}1  \\
\phantom{-}1  \\
-1 \\
-1  \\
\end{bmatrix} \right )
\end{align*}
where ``$\otimes$" denotes the Kronecker product. With this model matrix, we can assign rough interpretations to the $\beta_{g \ell}$'s in terms of log counts. For gene $g$, $\beta_{g 1}$ is the mean of the parent varieties B73 and Mo17, $\beta_{g 2}$ is half the difference between the mean of the hybrids and Mo17, $\beta_{g 3}$ is analogous for B73, and $\beta_{g 4}$ is half the difference between the hybrid varieties. Finally, $\beta_{g 5}$ is a gene-specific block effect that separates the first two libraries from the last two libraries within each genetic variety due to the samples being on different flow cells.

One goal of the original experiment was to detect heterosis genes: in the case of high-parent heterosis, genes with significantly higher expression in the hybrids relative to both parents, and in the case of low-parent heterosis, genes with significantly lower expression in the hybrids relative to both parents. 
For example, to detect genes with high-parent heterosis with respect to B73$\times$Mo17, we estimated $P\left (2\beta_{g 2} + \beta_{g4} > 0 \text{ and } 2\beta_{g 3} + \beta_{g4} > 0|y \right)$ using the cumulative mean technique described in Section \ref{PAPER1sec:gpu}.

We fit the model in Section \ref{PAPER1subsec:model} to the Paschold dataset using our CUDA-accelerated R package implementation, \fbseq{} and \fbseqCUDA{}. We used a single node of a computing cluster with a single NVIDIA K20 GPU, two 2.0 GHz 8-Core Intel E5 2650 processors, and 64 GB of memory. We ran 4 independent Markov chains with starting values overdispersed relative to the full joint posterior distribution. We ran each chain with $10^5$ iterations of burn-in and $10^5$ true iterations. We used a thinning interval of 20 iterations so that 5000 sets of parameter samples were saved for each chain. As in Section \ref{PAPER1sec:gpu}, we only saved parameter samples for the hyperparameters ($\tau, \nu, \theta_1, \ldots, \theta_L, \sigma_1, \ldots, \sigma_L$) and a small random subset of the gene-specific parameters. Running the 4 Markov chains in sequence, the total elapsed runtime was 3.89 hours.

To assess convergence, we combined the post-burn-in results of all 4 Markov chains. We used estimated posterior means and mean squares to calculate Gelman-Rubin potential scale reduction factors (see Section \ref{PAPER1sec:gpu}), which we used to monitor the $2L+2$ hyperparameters, the $G \times L$ model coefficient parameters $\beta_{g\ell}$, and the $G$ hierarchical variance parameters $\gamma_g$. All the corresponding Gelman factors fell below 1.1 except for $\beta_{26975, 2}$ (at 1.167), $\beta_{33272, 2}$ (at 1.148), $\gamma_{33272}$ (at 1.112), and $\beta_{6870, 2}$ (at 1.107). Although above 1.1, these last Gelman factors were still low and are not cause for serious concern. Next, for the total $2 \times 10^4$ saved parameter samples of each hyperparameter and of each of a small subset of gene-specific parameters, we computed effective sample size \citep{bda}. Most observed effective sizes were in the thousands and tens of thousands, the only exception being $\sigma_2^2$ at around 561 effective samples, well above the 10 to 100 effective samples recommended by \cite{bda}. There was no convincing evidence of lack of convergence.

\subsection{The scaling of performance with the size of the data} \label{PAPER1subsec:scale}

We used a simulation study to observe how the performance of our method scales with the number of genes and the number of \RNAseq{} samples. We used multiple new datasets, each constructed as follows. First, duplicate copies of the Paschold data were appended to produce a temporary dataset with the original 39656 genes and the desired number \RNAseq{} samples, $N$. Next, the desired number of genes, $G$, were sampled with replacement from the temporary dataset. We created 16 of these resampled datasets, each with a unique combination of $N = 16, 32, 48, 64$ and $G = 8192, 16384, 32768, 65536$ (i.e., $2^{13}, 2^{14}, 2^{15}$, and $2^{16}$, respectively). 

To each dataset, we applied the same method as in Section \ref{PAPER1sec:runtime}, with the same number of chains, iterations, and thinning interval. We also monitored convergence exactly as in Section \ref{PAPER1sec:runtime}. For 14 out of the 16 datasets, all Gelman factors of interest fell below our tolerance threshold of 1.1. For $G = 16384$ and $N = 16$, only the Gelman factors for $\beta_{9130, 3}$ (at 1.119) and $\beta_{13704, 2}$ (at 1.113) fell above 1.1. For $G = 65536$ and $N = 16$, there were 8 Gelman factors above 1.1. The highest of these was 1.325, and all corresponded to $\beta_{g\ell}$ and $\gamma_g$ parameters. Across all 16 datasets, the minimum effective sample size (ESS) for any hyperparameter was roughly 185 (for $\sigma_2^2$). Again, evidence of lack of convergence is unconvincing.

\begin{figure}[htbp]
   \centering
   \includegraphics[scale=0.75]{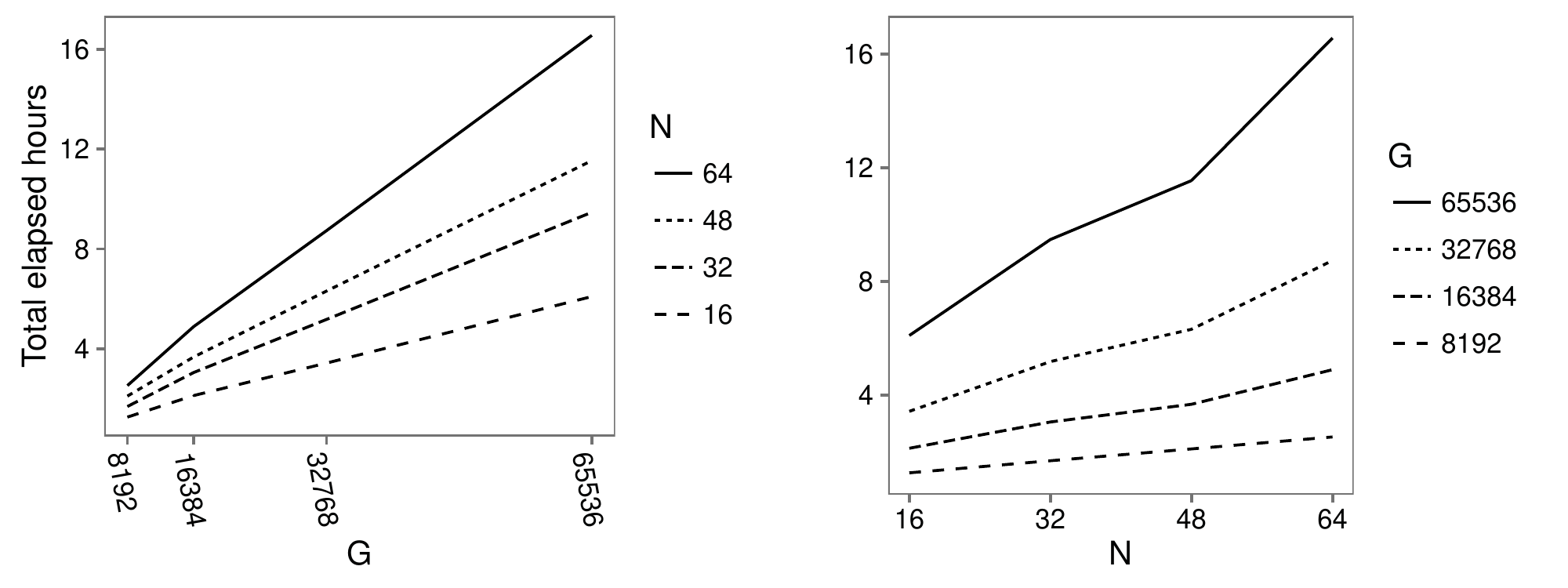}
   \caption{Elapsed runtime (hours) plotted against the number of genes ($G$) and the number of \RNAseq{} samples ($N$) for $2\times 10^5$ total MCMC iterations for four chains run in sequence.}
   \label{PAPER1fig:runtime}
\end{figure}

Figure \ref{PAPER1fig:runtime} shows the elapsed runtime in hours plotted against $G$ and $N$. Runtime appears linearly proportional to both $G$ and $N$ within the range of values considered. These runtimes, also listed in the runtime column of Table \ref{PAPER1tab:time}, vary from 1.27 hours to 16.56 hours. Our method appears expedient given the size of a typical \RNAseq{} dataset at the time of this publication. 

Table \ref{PAPER1tab:time} also shows ESS for hyperparameters ($\nu$, $\tau$, $\theta_1, \ldots, \theta_L$, $\sigma_1^2, \ldots, \sigma_L^2$). Overall, effective sample size appears acceptably high, and the time required to produce 1000 effective samples was relatively low for $N = 32$, 48, and 64. Of all the hyperparameters, $\sigma_2^2$ has the lowest ESS for $N = 16$. This parameter is the hierarchical variance of the $\beta_{g2}$ parameters, which, for the current model parameterization and on the natural log scale, are the gene-specific half-differences between the mean of all the B73xMo17 and Mo17xB73 expression levels and the mean of the B73 expression levels. For $N = 32$, $48$, and $64$, $\tau$ is the minimum-ESS hyperparameter. Recall that $\tau$ is the prior center (between the prior mean and the prior mode) of the $\gamma_g$ parameters, the counterparts of the gene-specific negative-binomial dispersion parameters often used in other models of \RNAseq{} data.

Naively, we should expect ESS to increase with both $G$ and $N$, since hyperparameter estimation generally improves with increased information to borrow across genes. Prior speculation about the time required to produce a given number of effective samples, however, is trickier. With additional data, estimation improves, but computation is slower. Table \ref{PAPER1tab:time} shows the interplay of these competing factors.

Many of the findings in Table \ref{PAPER1tab:time} are unsurprising given our prior expectations. Median ESS nearly doubled from $N = 16$ to $N = 32$ for all values of $G$ listed. Median ESS varied little among the larger values of $N$ and $G$, presumably since ESS is already close to the total aggregated $2 \times 10^4$ MCMC samples by that point. Next to median ESS in Table \ref{PAPER1tab:time} is the average time required to produce a median ESS of 1000 across the hyperparameters. For the larger values of $G$, there is a noticeable increase in this timespan between $N = 48$ and $N = 64$, and for $N$ fixed at 16, 32, 48 or 64, it increased roughly linearly with $G$. Minimum ESS, the minimum-ESS hyperparameter, and the time required to obtain 1000 effective samples of the minimum-ESS hyperparameter also showed some unsurprising trends. Minimum ESS increased from $N = 16$ to $N = 32$ for each value of $G$, and when $\tau$ was the minimum-ESS parameter, the time required to obtain 1000 effective samples increased roughly linearly with both $N$ and $G$. 

There are some surprises as well. Minimum ESS decreased with increasing $N$ when $\tau$ was the minimum-ESS hyperparameter and also decreases from $G = 32768$ to $G = 65536$ when $\sigma_2^2$ was the minimum-ESS hyperparameter. Also for $\sigma_2^2$ at $N = 16$, the time required to obtain 1000 effective samples decreased as $G$ increased from $8192$, to $32768$, but then spiked by the time $G = 65536$.

%The runtimes for $G = 8192$ and $N = 48$ deviate from this pattern the most, and this may be because those computations occurred several weeks after the rest. In the interim, the node was rebooted because of a power supply problem, and may have restarted in a different state.

\begin{table}[ht]
\centering
\topcaption{Runtimes and effective sample sizes for the simulation study in Section \ref{PAPER1subsec:scale}. $G$ is the number of genes, and $N$ is the number of libraries. The runtime column shows the total elapsed runtime in hours. The ESS columns show numerical summaries (either median or minimum, as indicated in the top row) of effective sample size across all hyperparameters ($\nu$, $\tau$, $\theta_1, \ldots, \theta_L$, $\sigma_1^2, \ldots, \sigma_L^2$). The ratio columns show 1000 times runtime divided by ESS: that is, the average elapsed hours required to produce 1000 effective samples (median or minimum across hyperparameters).}
\begin{tabular}{rrr|rr|rrl}
& & & \multicolumn{2}{c|}{median} & \multicolumn{3}{c}{minimum} \\ \hline
 G & N & runtime & ESS & ratio & ESS & ratio & parameter \\ 
  \hline
8192 &  16 & 1.27 & 10170 & 0.12 & 185 & 6.86 & $\sigma_2^2$ \\ 
  8192 &  32 & 1.69 & 19057 & 0.09 & 5473 & 0.31 & $\tau$ \\ 
  8192 &  48 & 2.11 & 18882 & 0.11 & 3810 & 0.55 & $\tau$ \\ 
  8192 &  64 & 2.53 & 19463 & 0.13 & 2303 & 1.10 & $\tau$ \\ 
   \hline
16384 &  16 & 2.13 & 10384 & 0.20 & 398 & 5.35 & $\sigma_2^2$ \\ 
  16384 &  32 & 3.05 & 18845 & 0.16 & 5018 & 0.61 & $\tau$ \\ 
  16384 &  48 & 3.68 & 19525 & 0.19 & 3590 & 1.02 & $\tau$ \\ 
  16384 &  64 & 4.89 & 19510 & 0.25 & 2663 & 1.84 & $\tau$ \\ 
   \hline
32768 &  16 & 3.43 & 12567 & 0.27 & 990 & 3.46 & $\sigma_2^2$ \\ 
  32768 &  32 & 5.18 & 18402 & 0.28 & 5501 & 0.94 & $\tau$ \\ 
  32768 &  48 & 6.31 & 19158 & 0.33 & 3279 & 1.92 & $\tau$ \\ 
  32768 &  64 & 8.73 & 19499 & 0.45 & 2450 & 3.56 & $\tau$ \\ 
   \hline
65536 &  16 & 6.09 & 11418 & 0.53 & 308 & 19.78 & $\sigma_2^2$ \\ 
  65536 &  32 & 9.47 & 18945 & 0.50 & 5554 & 1.71 & $\tau$ \\ 
  65536 &  48 & 11.54 & 19657 & 0.59 & 3624 & 3.18 & $\tau$ \\ 
  65536 &  64 & 16.56 & 19409 & 0.85 & 2673 & 6.19 & $\tau$
  \end{tabular}
\label{PAPER1tab:time}
\end{table}

\section{Discussion} \label{PAPER1sec:discussion}

We present a fully Bayesian strategy to fit large hierarchical models that are computationally demanding, possibly intractable, under normal circumstances. 
We introduce the two main components of most parallelized Markov chain Monte Carlo approaches: embarrassingly parallel computations and reductions. 
We combine these components with a slice-sampling-within-Gibbs MCMC algorithm, and we harness the multi-core capabilities of GPUs.
The CPU-GPU communication bottleneck is avoided by calculating running sums and sums of squares of relevant quantities. 
We demonstrate how these quantities can be used for convergence diagnostics and posterior inference.
We exemplified these general approaches using a real \RNAseq{} dataset and satisfied standard convergence diagnostics in 3.89 hours of elapsed runtime. 
In our simulation study based on the \RNAseq{} model, we found that total elapsed runtime scales linearly with the size of the data in each dimension within the range of sizes considered, and effective samples are hardest to obtain when the number of genes is high and the number of \RNAseq{} samples is low.

Major deterrents in the adoption of Bayesian methods are the development of computational machinery to estimate parameters in the model and the computation time required to estimate those parameters. 
General purpose Bayesian software such WinBUGS \citep{lunn2000winbugs}, OpenBUGS \citep{lunn2009bugs}, JAGS \citep{plummer2003jags}, Stan \citep{carpenter2016stan}, and NIMBLE \cite{nimble} have lowered the development time by allowing scientists to focus on model construction rather than computational details. 
These software platforms are based on DAGs representations of Bayesian models and  determine appropriate MCMC schemes based on these DAGs.
For analyses similar to the \RNAseq{} analysis presented here, estimation using these tools is far slower.
We hope the abstraction presented here and elsewhere, e.g.\ \cite{beam2015fast}, will inspire and spur the development of GPU-parallelized versions of these software enabling MCMC analyses of larger datasets and larger models.

%% file: acknowledgements.tex
% !TEX root = ../computation_paper.tex

\if0\blind 
{
\section{Acknowledgements} \label{PAPER1sec:ack}
\input{funding.tex}
} \fi

%% file: funding.tex
% !TEX root = ../thesis.tex
This research was supported by National Institute of General Medical Sciences (NIGMS) of the National Institutes of Health and the joint National Science Foundation / NIGMS Mathematical Biology Program under award number R01GM109458. The content is solely the responsibility of the authors and does not necessarily represent the official views of the National Institutes of Health or the National Science Foundation.

%% file: appendix.tex
% !TEX root = ../computation_paper.tex

\section{Univariate stepping-out slice sampler with tuning} \label{PAPER1appendix:slice}
\input{slice}

%% file: slice.tex
% !TEX root = ../computation_paper.tex

In the MCMC in Section \ref{PAPER1subsec:mcmc}, we repeatedly apply the univariate stepping-out slice sampler given by \cite{neal2003}. The goal of slice sampling is to sample $\theta$ from an arbitrary univariate density proportional to some function $f(\theta)$. To do this, Neal's method samples from $g(\theta, u)$, the bivariate uniform density on the region under $f(\theta)$ (i.e., $\{(\theta, u) : -\infty < \theta < \infty, 0 < u < f(\theta) \}$). The marginal density of $\theta$ under $g(\theta, u)$ is $f(\theta) / \int_{-\infty}^\infty f(\theta) d\theta$, so the samples of $\theta$ come from the correct target.

To sample from $g(\theta, u)$, Neal's method uses a technique similar to a 2-step Gibbs sampler. Here, suppose the current state is $(\theta, u) = (\theta^{(m)}, u_0)$. The first step of this two-step Gibbs sampler is to draw a new value $u\sim \Unif(0, f(\theta^{(m)}))$, the full conditional distribution of $u$ given $\theta = \theta^{(m)}$. The next step is to draw a new value $\theta^{(m+1)}$ of $\theta$ from the uniform distribution on $S = \{\theta: u < f(\theta)\}$, the conditional distribution of $\theta$ given $u$. Unfortunately, precisely determining the ``slice" $S$ is inefficient and not expedient in practice. %To compensate, the stepping-out slice sampler uses successive approximations of $S$ instead of $S$ exactly. 
The following explicit steps comprise a single stepping-out slice sampler iteration that moves from the current state $\theta = \theta^{(m)}$ to the next state $\theta = \theta^{(m+1)}$.

{\fontsize{10}{10}
\begin{algorithm}
\caption{Univariate stepping-out slice sampler with tuning}
\label{PAPER1alg:slice}
Set the initial size of the step $w$, total number MCMC iterations ($M$), number of burn-in iterations ($M_B$), number of initial iterations where $w$ is not tuned ($M_C<M_B$), maximum number of ``stepping out" steps ($K\in \mathbb{N}^+$), and $w_{\text{aux}}=0$. Let $\theta^{(m)}$ be the current value of $\theta$ at iteration $m$ of the MCMC chain. 
\begin{enumerate}
%\item Let $m$ $(m = 1, \ldots, M)$ be the index of the current MCMC iteration. Let $M_B < M$ be the number of iterations in the initial MCMC ``burn-in" period.
%\item Let $w$ be a tuning parameter. If $m = 1$, give $w$ some fixed positive scalar value. 
%\item Let $M_C < M_B$ be the number of burn-in iterations for which $w$ is not tuned.
%\item Let $w_{\text{aux}}$ be an auxiliary variable for tuning $w$. If $m = 1$, set $w_{\text{aux}} = 0$.
\item Sample $u\sim \Unif(0, f(\theta^{(m)}))$ distribution.
\item Randomly place an interval $(L, R)$ of width $w$ around $\theta^{(m)}$:
    \begin{enumerate}
    \item Sample $v\sim \Unif(0, $w$)$.
    \item Set $L = \theta^{(m)} - v$.
    \item Set $R = L + w$.
    \end{enumerate}
\item Set upper limits on the number of steps to perform in each direction:
    \begin{enumerate}
    \item Sample $K_L$ uniformly on $\{0, 1,\ldots,K\}$.
    \item Set $K_R$ = $K - K_L$. %$K_R$ is the maximum number of ``stepping out" operations to the right.
    \end{enumerate}
\item ``Step out'' the interval ($L$, $R$) to cover the ``slice" $S = \{\theta: u < f(\theta)\}$:
    \begin{enumerate}
    \item For $k=1,\ldots,K_L$, set $L=L-w$ if $u < f(L)$.
    \item For $k=1,\ldots,K_R$, set $R=R+w$ if $u < f(R)$.
    \end{enumerate}
\item \label{PAPER1itm:xstar} Sample $\theta^*\sim \Unif(L, R)$ distribution.
\item If $u < f(\theta^*)$, set $\theta^{(m+1)} = \theta^*$. Otherwise, set $R = \theta^*$ if $\theta^* > \theta^{(m)}$ or $L = \theta^*$ if $\theta^* \le \theta^{(m)}$, then go back to step \ref{PAPER1itm:xstar}.
\item \label{PAPER1itm:tune} If $m \le M_B$, tune $w$ as follows.
    \begin{enumerate}
    \item Increment $w_{\text{aux}}$ by $m \cdot |\theta^{(m+1)} - \theta^{(m)}|$.
    \item If $m > M_C$, set $w = w_{\text{aux}}/ (0.5 m (m + 1))$.
    \end{enumerate}
\end{enumerate}
\end{algorithm}
}

The tuning procedure in step \ref{PAPER1itm:tune} sets $w$ to be a weighted average of the absolute differences between successive values of $\theta$, giving precedence to later iterations. That way, $w$ is calibrated according to the width of the ``slice" $S = \{\theta : u < f(\theta)\}$. The popular black-box Gibbs sampler software, JAGS, uses this tuning method for its own slice sampler in version 4.0.1 \citep{plummer2003jags}.